\input harvmac

\input tables

\Title{hep-th/yymmnn RU-96-49}
{\vbox{\centerline{Quantum Hair on D-branes}
\centerline{ and Black Hole Information in String Theory}}}
\bigskip
\centerline{Tom Banks}
\smallskip
\centerline{\it Department of Physics and Astronomy}
\centerline{\it Rutgers University, Piscataway, NJ 08855-0849}

\bigskip
\baselineskip 18pt
\noindent

We introduce a notion of
quantum hair which completely characterizes the state of a D-brane in
perturbative string theory.  The hair manifests itself as a phase (more
generally a unitary matrix in subspaces of degenerate string
eigenstates) in the scattering amplitudes of elementary strings on the D
brane. As the separation of the D-brane and the string center of mass
becomes large, the phase goes to zero if we keep the string excitation
level fixed.  However, by letting the level number increase with the
distance, we can keep the phase constant.  We argue that this implies
that scattering experiments with highly excited strings can detect the
state of a D-brane long after it has \lq\lq fallen into a black
hole\rq\rq .
\Date{6/96}

\newsec{Quantum Hair and the Information Loss Paradox}

The notion of quantum hair was introduced some time ago by Krauss and
Wilczek\ref\kw{L.M.Krauss, F.Wilczek, {\it Phys. Rev. Lett.} {\bf 62}, (1986),380} as a proposal for resolving the information loss paradox in
Hawking radiation.  We will take it to mean the contention that every
quantum state of a black hole (and thus of anything that can be thrown into a
black hole) is characterized by some set of generalized Aharonov-Bohm
fluxes which enable an external observer to detect its presence by some
set of interference experiments performed outside the black hole.  These
fluxes are undetectable classically, and do not contradict the no hair
theorems. 

In its original context of internal Yang Mills gauge symmetries, it is
hard to know what to make of this idea.  It is hard to imagine a
symmetry large enough to characterize every state of the world.
Furthermore, Abelian Aharonov Bohm fluxes carried by localized objects
have a multiplicative character on asymptotic states. 
 That is, the flux carried by well separated objects 
is a product operator on the asymptotic Hilbert space.  If the flux
operators commute with the S matrix and indeed characterize each
individual state, then the S matrix will be unity.

In the present note we will show that string theory contains a notion of
quantum hair that avoids all of these objections.  We will demonstrate
this explicitly for D-branes, but by various string dualities this
implies that there is quantum hair for elementary string states as well.  
This adds to the growing list of evidence which indicates that string
theory avoids the information loss paradox by producing a unitary
S-matrix for black hole formation and evaporation\ref\lenny{L.Susskind,
{\it Phys. Rev.} {\bf D49}, (1994), 6606, hep-th/9308139; ``Some
Speculations About Black Hole Entropy in String Theory'',
hep-th/9309145;{\it Phys. Rev.} {\bf D52}, (1995),6997. } .  
We note that our proposal is somewhat reminiscent of the notion of
W-hair introduced in \ref\elsnwndrlnd{J.Ellis, N.E.Mavromatos,
D.V.Nanopoulos, {\it Phys. Lett.} {\bf B284}, (1992), 27.} .  
I do not understand
the precise relation between the two ideas, but I am fairly certain that
the conclusions I draw are different than those reached in that work.
J. Schwarz\ref\jhs{J. Schwarz, ``Target Space Duality and the Curse of
the Wormhole'', Presented at Oklahoma Conf. Beyond the Standard Model-
II, Norman, OK, Nov. 1-3, 1990.}
 has previously proposed that the infinite symmetry groups of
string theory might provide a source of quantum hair for black holes.
This work and a conversation with L. Susskind were the inspiration for
the current paper.  However, as Schwarz himself pointed out, the duality
symmetries which were the focus of his discussion are spontaneously
broken.  Thus, except at special points in moduli space where a tiny
piece of the gauge symmetry is restored, we do not expect them to
produce observable phases in the scattering from cosmic strings.
We will try to relate D-brane hair to string gauge symmetries, but
the gauge symmetries studied in this paper are more like large versions
of the continuous BRST symmetries of the perturbative string.  They are
not broken by the vacuum state.

\newsec{Hair on Branes}

We want to study the scattering of highly excited elementary strings
from D-branes in perturbative string theory.  For high excitation number,
the usual Fock basis for the string oscillator state is extremely
unwieldy. It is more convenient to remmber that strings are collections
of oscillators, and that the WKB approximation becomes exact for high
excitation energy.  Thus, to any periodic solution of the string
equations
\eqn\streq{(\partial_{\sigma}^2 - \partial_{\tau}^2 )X^{\mu} =0}
satisfying the classical Virasoro constraints, there corresponds a
Gaussian wave functional in the string Hilbert space which is
approximately BRST invariant.  As the excitation number increases, this
approximation becomes exact.  

Now consider the elastic 
scattering of one of these highly excited strings from
a D-brane.  It proceeds predominantly via the following mechanism.
If the classical configuration of the string intersects the D-brane
world volume, there is an (intrinsically quantum mechanical) amplitude
for the closed string to split into one or more open strings propagating
on the D-brane world volume.  For a highly excited essentially classical
string, these pieces will also be large classical objects.
 In leading order string perturbation
theory the open strings propagate freely on the world volume until they
reconnect to form the outgoing closed string.  

For scattering off the D-brane ground state, the open strings simply
have Dirichlet boundary conditions.  Excited states of the D-brane can
be accessed by adding a boundary operator $\delta L$, 
to the Lagrangian of the open
strings.  That is, excited states of the D-brane correspond to boundary
vertex operators (of course, for very large excitation of the D-brane this vertex
operator description will be very clumsy).  The amplitude will then pick
up a phase 
\eqn\phase{\delta = \sum \int \delta L}
where the sum runs over the trajectories of the open string ends on the
world volume.  This D-brane state dependent phase is the quantum hair.

The essential message here is that stressed in the penultimate section of 
\ref\banks{T.Banks, {\it Nucl. Phys.} {\bf B41}, (Proc. Suppl.),
(1995),21.}: the space of states of string theory simply does not split
up into a tensor product over a spacelike hypersurface.  States \lq\lq
localized at x\rq\rq , in the sense that their center of mass is in a
wave packet concentrated at $x$, are not independent of those at some
faraway point.

If, for a fixed string state, we let the distance between the D-brane
and the string center of mass go to infinity, then the classical
solution will no longer intersect the D-brane world volume, and the
phase $\delta$ will vanish.  However, {\it if we let the level number
increase with the distance, we will always find nontrivial phases}.
Thus, by doing scattering experiments with sufficiently highly excited
string states, we can in principle measure the state of distant
D-branes.  The root mean square distance from the center of mass of a
classical string of level  $N$  is $\sqrt{N} $.
Since the WKB approximation becomes exact for large excitation number,
this is also the average fluctuation of the string position in a typical
quantum string state of high excitation.

The last remark is the key to answering a possible objection to our
definition of quantum hair.  A physicist who believes strongly that
string theory is in some sense local might complain that our experiment for
measuring the state of a distant D-brane cannot actually be performed in
a laboratory far from the D-brane.  The classical string configuration
that we have described does intersect the D-brane.  How can we create it
by operations in our laboratory?  

To answer this objection, consider the amplitude for scattering $n+1$
 photons off the D-brane.  It has the form
\eqn\amp{\int_{Disk} V_{\gamma}(0) V_{\gamma}(z_1)\ldots V_{\gamma}(z_n)}
There is a region in moduli space in which all of the incoming photon
vertex operators are close to a point $z$ and all outgoing operators
close to another point $y$.  This is the region in which the
incoming photons form a highly excited single string state which then
scatters off the D-brane. 
Note that the question of finding a single highly excited state
in multiphoton scattering is independent of the boundary conditions on
the D-brane, because it is local in moduli space (we work at tree level).  

At tree level, the sum of the squares of the amplitudes for finding
strings of level $N$ in $k$ photon scattering is just the residue of the
pole in the $k\rightarrow k$ photon scattering amplitude, when the
incoming multiphoton invariant mass is $N$.  These residues are easily
calculated for two to two scattering and they do not go to zero with the
level (at generic values of the Mandelstam variable $t$).  Presumably it
is even easier to produce highly excited string states in multiparticle
collisions. 

Thus there seems to be no argument in principle against creating highly
excited string modes in collisions at sufficiently high energy.  Once we
have those modes available, our previous argument guarantees the
observability of the state of a faraway D-brane.  Of course, we will not
be able to pick the requisite phases directly out of a two photon plus D
brane to two photon plus D-brane amplitude, even though they in
principle contribute to the final result.  The scattering with a small
number of low level external states probes a complicated average of
excited state scattering amplitudes off the D-brane.  However, I believe
that scattering of large numbers of photons should provide us with
enough free parameters to probe individual phases.  This deserves
further study.  If it proves true, then the state of a faraway D-brane
will be encoded in complicated phase correlations in multiphoton
scattering amplitudes.  
The excitation level required for a
semiclassical string to pick up a state dependent phase in scattering
off a D-brane, increases like the square of the distance between the
D-brane position and the center of mass of the string.  We may expect a
corresponding increase with distance 
in the complexity (and in particular of the
number of photons) of a process which can measure these phases using
only external photons.

The application of these results to the black hole information problem
is straightforward.  Consider a large mass black hole in \lq\lq nice
slice\rq\rq coordinates.  These are coordinates which smoothly cross the
horizon, but asymptote to Schwarzchild coordinates at large spacelike
distances from the horizon.   It is in such coordinates that the black
hole information paradox is most sharply expressed in quantum field
theory.  For a black hole of very large mass and a region of spacetime
in which the Schwarzchild metric is approximated by the Rindler metric,
the spacelike distance between a supported and infalling observer grows
linearly in the nice slice time $T$.  Their relative velocity approaches
the speed of light as $1 - v^2 \sim {1\over T^2}$.  In field theory one
then argues that the supported observer cannot measure the state of the
infalling observer because no signal can ever propagate between them.
If however, the supported observer has at her disposal a highly excited
string (at her disposal means that the center of mass of the string
remains forever close to her spacetime position), this argument is no
longer valid, as we have seen above.  As time goes on, the observer must
perform a series of more and more intricate experiments at higher and
higher energies in order to extract information about things that fell
into the black hole.  These experiments are difficult beyond
comprehension, but perhaps not more so than those required to extract
the quantum information in the debris from an exploding hydrogen bomb.
By replacing {\it impossible} with {\it extravagantly difficult}, these
arguments return black hole information to the ordinary realm of
macroscopic phenomena.  

The stringy notion of quantum hair that we have outlined above does not
imply that the S-matrix is unity.  Although we concentrated on elastic
scattering above, there will also be processes in which the outgoing
closed string state is not the same as the incoming one.  For large
impact parameter, where we probe only very small momentum transfers, the
outgoing state must be degenerate in mass with the incoming state, but
the highly excited levels are also highly degenerate.  Thus the quantum
hair of a D-brane is not a collection of phases but rather a collection
of unitary matrices in the space of degenerate string levels.
These matrices will not commute with one another.  Furthermore, because
of the nonlocality of the hair, the matrix for two widely separated
D-branes will not be the product of the matrices for the individual
branes.   Thus, although the system has, in some sense, an infinite
number of conservation laws, they are not commuting, and do not act as
single particle operators in the space of asymptotic states.

Finally we note that various string duality transformations tell us that
elementary strings are really D-branes in another vacuum state of string
theory. D-brane hair can be measured in D-brane D-brane scattering as
easily as it can in the scattering of elementary strings.  Thus
elementary strings should have quantum hair.  We will get some hint of
its nature in the next section.

\newsec{D-brane Hair and Stringy Gauge Symmetries}

Quantum hair was originally discovered as the remnant of a spontaneously
broken gauge symmetry.  String theory is replete with such symmetries,
and it is not unreasonable to expect that D-brane hair can be related to
them. Indeed, the relationship is fairly straightforward.

Consider adding a total derivative to the lagrangian for a single
particle:
\eqn\partlag{\delta L = {d \over dt} F}
If $F$ depends only on the particle coordinate $X^{\mu}$
this has the form $\dot{X^{\mu}}\partial_{\mu} F$, which looks like the
response of a charged particle lagrangian to a gauge transformation.
In the case of a discrete internal gauge symmetry, there would be no
external potentials with nonvanishing field strength, but pure gauge
transformations of this type would measure Aharonov-Bohm flux and would
be the external evidence for quantum hair.
Of course, in the case of a particle we can only claim that this is so
after deriving the particle lagrangian from a theory in which there are
gauge fields {\it etc.}

By contrast, the rule of thumb in string theory is that {\it any} string
lagrangian which satisfies the conditions of conformal invariance
is an allowed classical background of the theory.  A lagrangian density
which is an exact two form $\delta {\cal L} = d \omega$, with $\omega $
a one form which is a smooth function of the string coordinates and
their derivatives, globally defined on target space, has no effect on
perturbative string scattering amplitudes.

To be more precise, we can think of 
the calculation of amplitudes in the presence of such a term in the
lagrangian as the computation of a functional integral with fixed
boundary conditions, $x^{\mu}_i (\sigma )$ (one boundary for each vertex
operator), followed by multiplication by functionals of the boundary
values representing BRST invariant states and integration over the
boundary values.  The total derivative will change the boundary path
integral by a factor $e^{i\int_B \omega}$, which will induce a 
unitary transformation on the space of BRST invariant states.
The important point is that it is the same transformation on each
external leg.  So there is no observable consequence of these
transformations. 

Consider however scattering off a D-brane of spatial dimension $p$.  
At tree level,
the amplitudes are computed
by computing vertex operator correlation functions in a superconformal field
theory with a single boundary with Dirichlet boundary conditions on 
$10 - p$ of the spatial coordinates.  This computes the scattering
amplitudes off the D-brane ground state.  Excited state amplitudes are
computed by adding a boundary vertex operator to the lagrangian.
This is a term $\int_{B_D} V_B$ where $V_B$ is a one form on the boundary of
the disk, constrained by boundary conformal invariance.  It is a
function of the bulk string coordinates, and so has an extension into
the interior of the disk.  The extension is not unique, since we can
add terms to V which vanish at the position of the D-brane.
Call the extension $V$.

Now write
\eqn\byparts{\int_{B_D} V_B = - \int_{D} dV - \int_{C_i} V}
where the first integral on the right is over the interior of the disk.
The $C_i$ are small circles surrounding the punctures where vertex
operators are inserted.   In general these will give nonzero
contributions to the amplitude, because the operator product expansion
of $V$ with the vertex operators will contain singularities.  The first
term on the right is a BRST exact operator $\{ Q, \{ b_{-1}, V\}\}$ and does
not contribute to amplitudes.\foot{The knowledgeable reader will have
recognized that we are merely repeating the derivation of axion
hair\ref\axion{M.Bowick,S.Giddings,J.Harvey, A.Strominger, {\it Phys.
Rev. Lett.} {\bf 61}, (1988),2823.} for more complicated stringy gauge
 transformations.}  

We began from a BRST invariant expression, and so the $\int_{C_i} V$ is
BRST invariant.  It's effect on the vertex operator $O_A$ inside $C_i$
is thus to replace it by a linear combination $H_A^B (V_{B}) O_B$ of the
complete set of BRST invariant vertex operators of the bulk theory.
$H$ is the \lq\lq trivial \rq\rq unitary operator we discussed above.

In essence, what we are saying above is that BRST transformations
which are trivial in the absence of D-branes are nontrivial when they
are present.  The simplest example is the diffeomorphism generated by a
vector field on target space.  This is clearly a trivial transformation
in the absence of D-branes, but it does not in general leave invariant
the Dirichlet boundary condition.  From this example it is clear that
all aspects of the D-brane state which are related to its configuration
in spacetime can be encoded in such nontrivial BRST transformations.
Our claim is that the same is true for any aspect of the D-brane state.

The scattering off different D
brane states can be related to the amplitude for scattering off any
reference state (above we chose the simple Dirichlet state) of the
D-brane multiplied by the $H$ matrices . Thus,
\eqn\Srel{S_{BB\ a_1 \ldots a_n} = H_{a_1}^{b_1}(B) \ldots H_{a_n}^{b_n}(B)
S_{DD\ b_1 \ldots b_n}}

Some of 
the phases are measurable.  Those $H$ matrices corresponding to
BRST transformations which leave the D-brane state invariant can still
be eliminated simultaneously from the amplitudes with and without a D
brane by performing the corresponding tranformation on both functional
integrals.  Those that do change the state of the D-brane constitute its
quantum hair.  The hair can be measured by for example preparing two
different D-brane states, scattering strings from them, and then
scattering the outgoing strings from the two experiments off of each other.

This argument does not by itself show that the $H$ matrices do not fall
off with distance, but from the previous section we know that this is
the case for highly excited external states.  More precisely, we must
increase the level of excitation with the distance in order to measure
stringy quantum hair on D-branes.  

The connection with \lq\lq nontrivial\rq\rq BRST transformations also
gives us a hint for the description of the quantum hair of elementary
strings. Elementary string states are in the BRST cohomology.  Thus, in
old fashioned language, they are gauge transformations with a gauge
function that is somehow singular.  The resemblance to Aharonov-Bohm
fluxes is striking.  In fact, Moore has obtained relations quite similar
to \Srel for open elementary string scattering 
by applying a set of symmetry transformations that are in one
to one correspondence with the BRST cohomology of the open
string\ref\moore{G.Moore, \lq\lq Symmetries of the Bosonic String
S-Matrix \rq\rq hep-th/9310026}. 
In general, Moore's transformations change the kinematic invariants in a
scattering process, but the lightlike Ward identities give relations
between amplitudes with zero momentum transfer.  It is possible in this
way to relate elastic scattering from any given string state to that
from any other. The relations I have found so far have
the general form of \Srel but
the analogs of the $H$ matrices are not unitary.  I believe that this is
a consequence of the fact that Moore studies infinitesimal
transformations, while I have discussed finite changes of state, but
this is an issue which deserves further study.  In particular, Moore's
formalism might enable us to establish the heuristic semiclassical
arguments of the present paper in a much more rigorous fashion.

Finally, it is worth pointing out that the formula  \Srel encodes the
information about D brane states in leg pole factors, reminiscent of
those of $1 + 1$ dimensional string theory.  This is the advertised
connection with the work of  \elsnwndrlnd .  Again, this suggests a
direction for further research on the vexing problem of the formation of
black holes in $1 + 1$ dimensions.

\newsec{Conclusions}

We have argued that perturbative string theory contains a notion of
quantum hair for D-brane states.  The hair takes the form, for the most
part, of Berry matrices in the scattering amplitudes of highly excited
degenerate string states from the D-brane. 
We have argued that the quantum hair carries complete information
about the D-brane state, so no information about the D-brane is lost as
it falls into a black hole.  Since perturbative string states are
simply D-strings of another classical 
string vacuum state, all of these statements
must be true of general states in string theory.

There are a large number of points where the argument has been less than
rigorous. Our derivations have been highly semiclassical and have
neglected subtleties of ghosts and BRST invariance that may well be
important to a proper understanding of quantum hair.  The work of Moore
provides the most likely entry into a fully quantum derivation of these
results.  It is also likely to be the correct description of quantum
hair for elementary string states.  Finally, all of our work has relied
on perturbation theory and the assumption that highly excited string
states were stable.  One would like to understand how to go
beyond such tree level analysis, but the treatment of unstable particles
even in weakly coupled string theory is at the moment beyond our
capabilities. 
\centerline{\bf Acknowledgements}

The genesis of this paper was a conversation with L.Susskind about quantum
hair in string theory in the spring of 1995.  
I also owe him a debt of gratitude for many patient
explanations of his groundbreaking ideas about the resolution of the
black hole information paradox.   Steve Shenker independently
thought of using D-branes as a local probe of black hole dynamics, and I
would like to thank him for a number of conversations about high energy
scattering in string theory. 
Mike Douglas\ref\mrd{M.R. Douglas, ``Gauge Fields and D-branes'', hep-th/9604198. } has also emphasized the idea of D-branes as
local probes. G.Moore was kind enough to guide me through his beautiful
results on bosonic string Ward identities. Finally I
would like to thank J.Polchinski
and D.Friedan for important conversations about boundary operators and BRST
invariance, and Joel Shapiro for a discussion of the production of
highly excited string states.  
This work was supported in part by the Department of Energy
under grant $\# DE-FG02-96ER40959$
\listrefs
\end